\documentclass[a4paper,UKenglish,cleveref, autoref, thm-restate]{clean-lipics-v2021}
\title{Quick(er)sort\thanks{Dedicated to the memory of
 C.A.R. Hoare} and optimal adaptivity} %TODO Please add

\author{Sandeep Sen }{IIIT Delhi , India }
{ssen@iiitd.ac.in}{}{}
\authorrunning{S. Sen} 
\Copyright{Sandeep Sen} 
 %TODO mandatory: Please choose ACM 2012 classifications from https://dl.acm.org/ccs/ccs_flat.cfm 

 %TODO mandatory; please add comma-separated list of keywords

\newcommand{\ignore}[1]{{}}

\newcommand{\E}{{\mathbb E}}
\usepackage{epsfig,fancyhdr,amsmath,amssymb,latexsym}
\usepackage[ruled,vlined]{algorithm2e}
\usepackage{graphicx,xcolor,url}
\usepackage{colortbl}
\hideLIPIcs
\date{}
\nolinenumbers
\begin{document}
%\title{ Quicksort and universal optimality}
%\author{Sandeep Sen\thanks{Email:\texttt{sandeep.sen@ashoka.edu.in}}\\
%Computer Science Department\\ Ashoka University,
%India}
\maketitle
\begin{abstract}
We use a robust  measure of presortedness, based on the notion of
$\alpha$-sorted sub-sequences, to
show that Quicksort, in general, Partition sort, is optimal for a much broader
class of inputs than those captured by worst-case analysis.
Our result also generalizes the well-known result on sorting multisets with 
repeated elements (called multisorting) 
to a more general class of permutations that is based on entropy measure.
We revisit some existing sorting algorithms from this perspective, and 
show that unlike previous entropy based measures, we can prove a nearly 
matching information theoretic lower bound. 
%incorporating more explicit and implicit parameters into the 
%analysis can lead 
%us to more general notions of efficiency called universal optimality.
%This framework captures input-sensitive behavior in a unified manner and 
%suggests a principled approach to analyzing algorithms beyond 
%traditional worst-case or average-case paradigms
This perspective can open up interesting directions for adaptive 
design and analysis of fundamental algorithmic problems beyond 
traditional worst-case or average-case paradigms.
\end{abstract}
%\newpage
\section{Introduction}

 Sorting is one of the most extensively studied problems in Computer Science 
with a rich body of work spanning more than five decades.
In the 
comparison model, optimal $O(n\log n)$ algorithms have been discovered as early as
the 1960s, although in the RAM model, it is still not fully resolved. 
Knuth \cite{Knuth3}
dedicated an entire volume to various aspects of this fundamental problem. 
The worst-case (as well as average-case) lower bound follows from the 
information-theoretic argument of $\log ( n ! ) = \Omega ( n \log n )$ 
where $n !$ is the 
number of permutations of a given set of $n$ elements. 
Franceschini and Geffert 
\cite{FG:05} could achieve optimality in the number of data movements along with 
optimal
space (in-place) and time. 
We will confine ourselves
to the comparison model and address the following important question: 
how to design algorithms that take advantage of existing presortedness in
the input sequence and perform better than worst-case guarantees.
%We want to consider a variation of the worst-case optimal
%algorithms, viz., if we had
%more information about the input permutation, can we take advantage of that, instead
%of considering the entire space of $n!$ permutations ?

In this work, we introduce a new measure of presortedness based on 
$\alpha$-sorted subsequences. Informally, these are subsequences that appear 
in sorted order and correspond to contiguous blocks in the final sorted 
sequence. This notion generalizes multisorting by replacing repeated values 
with sorted contiguous blocks of the final sorted permutation. 

Consider a multiset that has $n_i$ elements with value $v_i$, $1 \leq i \leq k$ 
where $\sum_i n_i = n$. 
This can be sorted in $O(n \log k)$ time by maintaining a balanced tree
by storing only the $k$ distinct elements - sometimes referred to as tree-sort. 
A running time of  $O(n \cdot H 
( \frac{n_1}{n},\frac{n_2}{n},\ldots \frac{n_k}{n} ))
+ O(n)$ can be achieved \cite{MS:76,MR:91} where, 
$H ( x_1 , x_2 , \ldots x_n) $ denotes the entropy function
$\sum_{i=1}^n x_i \log (\frac{1}{x_i})$ where $0 \leq x_i \leq 1$ and 
$\sum_{i=1}^n x_i = 1$. Since this function is crucial in the context of this
paper, it is straightforward to verify that $O( n \cdot H 
( \frac{n_1}{n},\frac{n_2}{n},\ldots \frac{n_k}{n} ))$ is always
bounded by $O(n\log k)$ and this bound is also known to be optimal for the 
multisorting problem (\cite{MS:76}). We generalize this result, 
where, instead
of $n_i$ elements with identical values, we have sorted subsequences of
lengths $n_i$.  For example, the input [ 12, 4, 6, 4, 12, 4, 6, 4] will get 
sorted as 
[4, 4, 4, 4, 6, 6, 12, 12], using any multisorting algorithm 
where 4, 6, 12, have repeated occurrences, viz., four, two and two respectively. 
In contrast, for a sequence like
[ 12, 4.1, 6.2, 4.3, 12.7, 4.5, 6.5, 4.8], then the multisorting algorithms 
becomes ineffective, since 
we have increasing subsequences like [4.1, 4.3, 4.5, 4.8] and [6.2, 6.5]  etc.
instead of repeated values.
The precise definition of such sequences will be given in the next section.
\subsection{Related work}
There is an existing body of work that studies this issue using different 
measures of pre-sortedness like number of inversions or maximum displacement. 
For example, we know that a (nearly) sorted
input instance could take $O(n)$ comparisons as opposed to the worst case bound
of $\Theta (n \log n)$. 
In general, if all elements were within distance $1 \leq 
k \leq n-1$ of their final sorted positions, it is known that such 
sequences can be sorted
in $O(n k)$ steps using $k$ iterations of insertion sort. Using a heap of
size $k$, this can be done more efficiently in $O(n\log k)$ time.
Some algorithms \cite{EF:03,SS:03} based on heapsort can adapt to the number
of inversions $I$
in the input (a pair $( x_i , x_j )$ is {\em inverted} if $i < j$ but 
$x_j < x_i$)
and achieve a running time of $O( n \log (I/n + 1))$. 

\begin{figure}[h]
\begin{center}
\begin{tabular}{|c|c|c|c|c|c|c|c|c|c|c|c|c|c|c|c|}
\hline
4 & 6 & 8& 23 & 25 & 33 & 34 & 39 & 45 & 55 & 56 & 62 & 68 &
72 & 84 & 85 \\
\hline
\rowcolor{yellow}
6 & 4 & 23 & 8 & 33 & 25 & 39 & 34 & 45 & 56 & 55& 62 &
72 & 68 & 85 & 84 \\
\hline
\rowcolor{pink}
45 & 55 & 56 & 62 & 68 & 72 & 84 & 85 & 4 &
 6 & 8 & 23 & 25 & 33 & 34 & 39 \\
\hline
4 & 8&  25 &  34 & 45 &  56 & 68 & 84 & 6 & 23 &33 & 39 &55&62 &72 &85 \\
\hline
\end{tabular}
\end{center}
\caption{The 2nd row has elements within distance 1 and 3rd and 4th rows within
distance 8 or $n/2$ for a row with $n$.
Further, the number of inversions in 2nd, 3rd, and 4th rows are $n , n^2 /2$ 
and $n^2 /2$
respectively.\\
However, there is still a large degree of sortedness in 3rd and 4th row.} 
\label{unsorted}
\end{figure}
However, these measures fail to capture certain forms of structured order 
that arise naturally in practice.
Consider, for instance, a sequence formed by concatenating large sorted 
blocks or by interleaving sorted segments. Such inputs may contain a large
number of inversions or large displacements, yet possess significant 
natural ordering. Traditional measures do not adequately reflect this 
structure, and consequently do not enable algorithms to exploit it effectively.
Figure \ref{unsorted} illustrates 
some interesting input permutations.

There is an extensive literature (\cite{VW92,PM95,MW18}) 
that has tried to quantify pre-sortedness of the input
beyond the displacement and inversion measures.
Following an earlier work of
Mannila \cite{Ma:85}, Petersson and Moffat \cite{PM95} provide an elaborate 
account of different measures of
sortedness including relationship between these measures. 
A particularly influential approach pioneered by \cite{LP:94} is 
to identify sorted 
runs (sorted subsequences)
referred to as SUS (Shuffled-up Sequences) or SMS (Shuffled Monotone  
Sequences) that includes both increasing and decreasing runs. These are merged 
using some versions of 
classical optimal merging algorithms based on Huffman-tree. 
A popular and successful implementation of this family of
 {\em adaptive mergesort algorithms} 
is known as {\it Timsort} that uses runs of consecutive elements, 
is a part of Java and Python library.
Using sophisticated techniques, 
\cite{TN10,BN13,GJKY:25} obtain the bound 
of $n \cdot H ( \frac{n_1}{n},\frac{n_2}{n},\ldots \frac{n_k}{n} ) 
+ O(n)$. 
%$H ( x_1 , x_2 \ldots x_n) $ denotes the entropy function. 
%$\sum_{i=1}^n x_i \log (\frac{1}{x_i})$ where $0 \leq x_i \leq 1$ and $\sum_{i=1}^n x_i = 1$. 
Optimality is difficult to establish 
because the running time depends on the lengths of the sorted runs in SUS/SMS
and these may not be the best possible decomposition of the input sequence 
(the intractability of this problem was shown in \cite{KSW96}). 
%Based on comparison-optimal merging schemes like exponential merging 
%\cite{Mc:93}, the paper
%\cite{GJKY:25} shows how to merge runs of sorted sequences with tight bounds
%on the comparisons.   

\subsection{Main contributions}

We first present a measure of sortedness that views the input as a union 
of $\alpha$-length 
sorted sub-sequences. This measure is closely related to the notion of 
{\it dual runs} defined in \cite{GJKY:25} but we consider more general cases
that involve subsequences with non-constant values requiring more careful
treatment. We 
show that this measure is better
suited to (a simple variation of) {\em partition sort} compared to 
sophisticated versions of mergesort. 

More specifically, we prove that, if the 
input consists of sorted subsequences of sizes ($n_1, n_2, \ldots, n_k$), 
then the algorithm achieves a running time of
\[
O\left(\sum_{i=1}^k n_i \log\left(\frac{n}{n_i} + 1\right)\right)
= O\bigl(n \cdot H(\frac{n_1}{n}, \frac{n_2}{n}, \ldots \frac{n_k}{n}) + n\bigr).
\]
Note that this decomposition is not known to the algorithm.
We also establish lower bounds, showing that this performance is 
optimal for this entire class of inputs except for $k \in n - o(n)$, 
the precise bounds are stated in Theorem \ref{lbnd} and Lemma \ref{long-seq}. 
The significance of this result using our measure of sortedness is illustrated
by the following example, where we compare this with the
distance-based and the inversion-based  measures in Figure
\ref{unsorted}. The input in row 3 has $\Omega ( n^2 )$ inversions and a
distance bound of $\Omega (n)$ that could lead to $\Omega (n \log n)$ bound
for both the distance-adaptive and the inversion-adaptive algorithm. However,
using the measure of $\alpha$-sorted subsequences, 
the algorithm for $( n/2 , n/2)$, runs in $O(n)$ time.
Further, we show that randomized Quicksort 
augmented with stable partitioning, that we call 
{\it Quickersort}, has an
expected running time that matches the aforementioned bound
(Theorem \ref{randoptsort}). 

The connection to multisorting is interesting in
its own right and the results in this paper may be seen as a non-trivial 
generalization of the known results of {\it multisorting}.  
One of our goals is to motivate further research in
a more general framework of algorithm design
for which sorting can provide important insights. 

Existing entropy-based adaptive sorting bounds rely on decompositions defined 
on the basis of the input (e.g., runs or subsequences). In contrast, our 
decomposition is defined with respect to the final sorted permutation, which
enables us to obtain a stronger entropy characterization of the complexity 
of sorting. To the best of our knowledge, there is no prior work that achieves
provable entropy-based optimal results, except multisorting. 

Our results suggest a broader perspective on algorithmic efficiency.
%we call \emph{universal adaptivity}. 
Rather than analyzing performance 
with respect to a fixed input distribution or a single parameter, 
this framework evaluates the effectiveness of 
algorithms relative to the complexity for each 
sub-class of inputs. In the context of this paper, this is given by the
vector $( n_1 , n_2 \ldots n_k )$. 
Sorting serves as a natural testbed for this idea, but we 
believe the underlying principles extend to other domains where input 
structure plays a crucial role. Similar bounds had been obtained for the 
problem of convex hull and the dominance problem in \cite{SenG99,ABC17}.

The remainder of the paper develops this framework in detail. We formally 
define $\alpha$-sorted subsequences, present the partition-based algorithm and 
its 
analysis, derive lower bounds, and then re-analyze randomized Quicksort for
this variation.
%It is worth noting that the running time of the algorithm is
%also inherently related to the output and a common example is that of computing the
%$n$-th Fibonacci number. While the input has size $\log n$ bits, the output is
%$\Omega ( n)$ and any efficient algorithm would compute this in $O( \log n + h )$
%steps where $h = n$ is the size of the output. This is not very interesting as the
%size of output is known in advance. 

\section{A robust characterization of sortedness} 
We consider a natural measure of sortedness that is not known explicitly from
the input, and design an algorithm that exploits this measure.  

{\bf Definition } ($\alpha$-sorted subsequence): For a given input $S$, 
and $i_1 < i_2 < \ldots i_{\alpha}$, let
$x_{i_1} \leq x_{i_2} \leq \ldots x_{i_{\alpha}}, \ \ \ x_{i_j} \leq 
x_{i_{j+1}}$ be a sorted subsequence.
Consider the sorted permutation $\sigma (S)$. If
 $ x_{\sigma( i_1 )},
x_{\sigma( i_2 )} , x_{\sigma( i_\alpha )}$ are consecutive in the
sorted permutation $\sigma(S)$, 
i.e., 
for all $j$, $\sigma( i_j ) = \sigma ( i_{j-1}) +1$, then this subsequence is 
called an $\alpha$-sorted subsequence.   

\begin{remark}In the remaining paper, we will consider only stable sorted 
permutation, where repeated
values appear in the same relative order as their input ordering.
This requirement differs from the traditional multisorting, where arbitrary
orderings of equal-valued elements simplify the analysis. 
\end{remark}

We call an $\alpha$-sorted subsequence {\it maximal} if it is not 
contained in a longer $\alpha'$-sorted subsequence where $\alpha' > \alpha$. 
See Figure \ref{alphadef} for an illustration.\footnote{
An alternate equivalent characterization is provided in the proof of 
Lemma \ref{long-seq} and some readers may find it useful.} 
\begin{lemma}
The maximal $\alpha$-sorted subsequences are disjoint.
\label{disj}
\end{lemma}
\begin{proof}
Suppose $y \in A \cap B$ where $A, B$ are two non-identical $\alpha$ 
subsequences. Since $\sigma ( A)$ and $\sigma (B)$ are rank consecutive in the
sorted set, the predecessor (successor) of $y$ would also belong to both 
subsequences because of the maximality assumption. 
Therefore $A, B$  must be identical, contradicting the assumption that 
proves the statement.
\end{proof}
Consequently, this implies that the decomposition of a given input into maximal 
$\alpha$-sorted subsequences is unique, and will be referred to as a canonical
decomposition.

Let an input
contain maximal sorted subsequences 
$\{ \alpha_1 , \alpha_2,  \ldots \alpha_k \}$ 
where $| \alpha_1 | = n_i$ and
$\sum_{i=1}^k n_i = n$.
See Figure \ref{alphadef} for an illustration.
Explicitly determining the $\alpha_i$'s in a given input is straightforward if 
we know the sorted sequence but not otherwise. 
Further, we will not distinguish between the permutations of $\alpha_i$'s and
refer to them as the same {\it sorted-type}.
\\
\begin{figure}[h]
\begin{tabular}{|c|c|c|c|c|c|c|c|c|c|c|c|c|c|c|c|}
\hline
4 & 6 & 8& 23 & 25 & 33 & 34 & 39 & 45 & 55 & 56 & 62 & 68 &
72 & 84 & 85 \\
\hline
62 & \cellcolor{pink} 23 & 6 & \cellcolor{pink}  25 & 85 & \cellcolor{pink} 33 & 8 &
\cellcolor{pink} 34 & \cellcolor{pink} 39
 & 84 & 72 & 55 & 56 & 4 & \cellcolor{yellow} 45 &   68 \\
\hline
\end{tabular}
%\\
\caption{In the example above, 23, 25, 33, 34, 39 is a 5-sorted, but not maximal, as
45 can be appended.\newline
The above sequence is a 2-6-2-1-1-1-2-1 sequence
corresponding to (62,68) (23, 25, 33, 34, 39, 45) (6,8) (85)(84)(72)(55,56)(4)
sub-sequences. }
\label{alphadef}
\end{figure}

%\fbox{\parbox{\textwidth}
   %{
  This definition of $\alpha$-sorted sequence is a generalization of the
notion of {\it dual runs} of the inverse permutation in \cite{GJKY:25} 
who in turn is related to the definition of {\it riffle shuffles} in 
\cite{Mc:93}. The riffle shuffles are permutations resulting from partitioning
the sorted sequence into $s$ intervals and permuting the elements 
while maintaining the relative ordering of the elements within each interval.
Although the notions appear identical, there is a subtle difference when 
the inpputs is not restricted to $s \leq n$ distinct values.
This induces at most
$ s$ intervals where a dual run corresponds to all the elements with 
identical values. It is easy to see that the
dual runs corresponding to a fixed value could be a 
sub-interval of the more general case 
of $\alpha$-sorted subsequences. This is
because of additional elements whose values are strictly smaller or larger.
The definition in 
\cite{Mc:93} referred to {\it monotonic} sub-intervals 
but didn't require maximality of a monotonic runs since it included all the
elements with the same values. 
We need a more careful definition that includes maximality
as illustrated in the previous example and it is
critical for most results in this paper. It forces an algorithm 
to exploit a lower entropy bound; moreover  our algorithm 
adapts to this generalized notion of partial sortedness without the
knowledge of the $\alpha_i$'s.\footnote{ 
In \cite{GJKY:25}, the authors exploited
the technique of adaptive exponential merging in conjunction with dual runs 
to obtain optimal bounds 
for inputs with $\leq s$ distinct values which is much restrictive
than the generalized entropy based bounds in this paper.} 
%}
%} 
%\\[0.1in]

In this paper, we show that this class of permutations can
be more effectively dealt with using  
a variation of the generic {\it partition sort} (Figure \ref{partitionsort}). 
We use the term {\it partition sort} to refer to any algorithm that uses a 
{\it pivot} to
partition the set into two subsets - one where the elements are smaller 
than the pivot
and the other where they are greater than the pivot.
Subsequently, this strategy is applied recursively to each of the partitions. 
The running time is sensitive to the relative sizes of the two partitions.
A simple strategy that is used by Quicksort, 
is to choose a random element as a pivot, 
whereas we can use a median 
to split the input into exactly equal size halves. Recall that one can find
a median in deterministic linear time. The partitioning used by Quicksort is
harder to analyze which will be address later.   

Once the pivot is known, the actual partitioning can be done easily in linear
time. Here, we will require our partitions to satisfy an additional property. 
\begin{remark} A {\it stable partitioning} procedure has the property that for 
any pair of
elements $( x_i , x_j )$, $i < j$, they will be in positions $i' , j'$ 
respectively such 
that $i' < j'$ after the partitioning if they belong to the same partition.
\end{remark}
See Figure \ref{stablepart} regarding the significance of this. Although
 stable partitioning
can be done in $O(n)$ time easily; however, an in-place stable partitioning 
in linear time is considerably more complex \cite{KP:92}.

\begin{figure}[h]
\begin{tabular}{|c|c|c|c|c|c|c|c|c|c|c|c|c|c|c|c|}
\hline
\rowcolor{pink}
45 & 55 & 56 & 62 & 68 & 72 & 84 & 85 & 4 &
 6 & 8 & 23 & 25 & 33 & 34 & 39 \\
\hline
4 & 6 & 8& 23 & 25 & {\bf 33} & 45 & 55 & 56 & 62 & 68 &
72 & 84 & 85 & 34 & 39 \\
\rowcolor{yellow}
23 & 6 & 4 & 25 & 8  &{\bf 33}  & 85  & 34  &72  & 62 & 55 &56  &
68 &84  &39  & 45  \\
\hline
\end{tabular}
\caption{ Row 2 is a stable partitioning around 33 but row 3 is not. Note that stable partition
around 39 would yield a sorted set}
\label{stablepart}
\end{figure}
 We provide a simple application of stable partitioning. 
Consider an input where
the first $n/2$ elements
are the sorted second half of the array and the second $n/2$ elements form the 
sorted first half (Figure \ref{unsorted}, row 3). In this case, $k = n/2$, 
however, if we partition the input around the median, ensuring {\it stability}, 
we arrive at a sorted array which is much better than 
the $O(n \log k )$ bound. Therefore, 
the definition of {\it sortedness} based on distance 
appears weak for this situation.
%A sorted array has zero inversion whereas a reverse sorted array has ${ n \choose 2 }$
%inversions. The previous example also has 
%$\Omega ( n^2 )$ inversions and so this
%measure of sortedness also fails to take advantage of ordering available in the array.

\begin{figure}[h] 
\begin{procedure}[H] 
 \nl {\bf Input:} An unordered array $A$ of $n$ numbers $x_1 , x_2 \ldots x_n$  \;
 \nl \If {$A$ is not sorted (verify by comparing adjacent elements)}{
\nl Use the median element $M$ as pivot, and partition $A$ around $M$ using a stable procedure \;
 Stable-Partition-Sort ( $A_< $) /* $A_<$ : elements smaller than $M$ */ \;
 Stable-Partition-Sort ( $A_> $) /* $A_<$ : elements greater than $M$ */ \;
}
\nl {\bf Output} $A$ 
\caption{Stable-Partition-Sort($A$) }
\end{procedure}
\caption{ Stable version of Partition Sort}
\label{partitionsort}
\end{figure}
Note that we are using a minor modification of partition sort where 
we verify if the
input is sorted before any recursive call. 
Since this requires only a linear time
check, it doesn't increase the asymptotic cost of partition sort. 
Moreover, our partitioning procedure 
should preserve the relative ordering of the
elements in each half as per the input ordering. 
\begin{theorem}
An input consisting of $( n_1 , n_2 \ldots n_k )$-sorted sub-sequences 
can be sorted using\\
$O( \sum_{i=1}^{k} n_i \log ( \frac{n}{n_i} + 1 ) ) = O( n \cdot H ( \frac{n_1}{n} , 
\frac{n_2}{n} \dots \frac{n_k}{n}) + n)$ comparisons where $\sum_i n_i = n$.
\\
Moreover the algorithm doesn't require prior knowledge of the vakues $n_i$.
\label{optsort}
\end{theorem}
We will first prove a simple property that forms the crux of our analysis.
The choice of medians as a partitioning elements corresponds naturally to a family
of dyadic intervals denoted by 
\[ I_{d,j} , \ \ d = \{ 0, 1 \ldots \log n \} , \ 
j = \{ 0 , 1 \ldots  2^d \}  \]
where $d$ denotes the level of recursion and $j$ corresponds to the labels of
the $ 2^d $ intervals of sizes $\frac{n}{ 2^d }$ in level $d$. 
These intervals are induced by the consecutive medians chosen at level $d$ that 
are denoted by $M^d_j \ j = 0 , 1 \ldots 2^d$. The number of elements in the
interval $[ M^d_j , M^d_{j+1}]$ equals $\frac{n}{2^d}$. Note that the intervals
are not of the same geometric length but the spans that they cover, suffices 
  for our purpose. 

The 
proof is based on a fundamental property of partitioning of an arbitrary interval
of size $m \in [1, n]$ that corresponds to the length of any  $\alpha$-sorted 
subsequence. For simplicity, assume that $n$ is a power of 2. 

Since the elements of ${\cal S}$ are consecutive in the sorted
output. So any interval greater than $2^p$ must contain a pair of elements that are
separated by $2^{p-1}$. Figure \ref{consec} illustrates this property.
There cannot be consecutive such pairs in level $\log n - (p +1)$
since $| {\cal S} | < 2^{p+1}$; however, there may be consecutive pairs from level
$\log n -p$, that will not be required in the analysis.
\begin{figure}[h]
\includegraphics[width=5in]{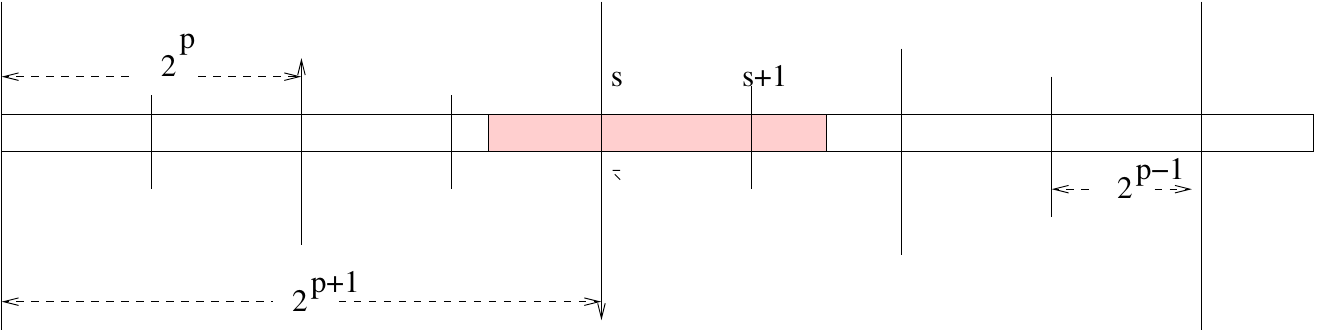}
\caption{The figure shows overlays of partitions of sizes
$2^{p+1} , 2^p , 2^{p-1}$. The shaded interval ${\cal S}$ has length $L$ where
$2^{p+1} > L > 2^p$ and there are two consecutive
elements $[s, s+1]$
of partition $2^{p-1}$
that must fall within ${\cal S}$.
} 
\label{consec}
\end{figure} 

It is known that any interval of length $m , \ 2^{p} \geq 
m < 2^{p+1} $ can be covered by a set of at most $2p $ 
intervals of the family of dyadic intervals described above. Moreover, these
are of sizes at most $2^p$.
This follows from a well-known result on the
decomposition of any given interval 
induced by a {\it segment tree} (see \cite{PS:85} for details). An 
interval ${\cal S}$ splits into sub-intervals at some level of the recursion, where it straddles the pivot.
Each sub-interval can be seen as a semi-interval whose length is a union of
dyadic intervals and this is just like a binary representation that implies
the above property.

%In the sorted set the $m$ elements are contiguous and
%so are all the dyadic sub-intervals. As the partitioning gets refined
%recursively the corresponding sub-intervals do not require further recursive 
%calls because of the stable partitioning property. Our analysis involves a
%charging argument for the cost of partitioning of the $m$ elements till the 
%recursion terminates at the levels of the sorted (dyadic) sub-intervals.

%Let us denote the subset of partitioning elements
%in level $r$ (starting with $r=0$ at the top) by 
%$M^r_s, \ \ s = 0, 1 \ldots$ where $M^r_i ,
%M^r_{i+1}$ are consecutive partitioning elements in level $r$. The partitions are of sizes
%$2^{\log n - r}$ assuming $n$ is a power of 2. 
The above discussion can be formalized as follows.
\begin{lemma}
Consider a maximal $\alpha$-sorted sub-sequence ${\cal S}$ such that
$|{\cal S}| = m$. Then the ranks of the elements of ${\cal S}$ form a 
contiguous interval of size $m$ in the sorted set that can be decomposed
into a union of canonical intervals of $I_{d,j}$, 
such that there are at most two intervals in the same level $d$.
So ${\cal S} = \bigcup_{r=1}^{r=t} B_r$  where $B_r \in I_{d,j}$ where
at most two intervals $B_1 , B_2 \ldots B_t$  belong to any fixed level $d$.
\end{lemma}
\begin{proof}(of Theorem \ref{optsort})
For simplicity, assume first that $n$ is a power of two and that every
partition is performed using an exact median. Thus, at recursion depth $d$,
the sorted order is partitioned into $2^d$ consecutive canonical intervals,
each of size $n/2^d$. Equivalently, these are the dyadic intervals associated
with the balanced partition tree.
Fix a maximal $\alpha$-sorted subsequence ${\cal S}$ of length
$m = |{\cal S}|$.

Since the elements of ${\cal S}$ are consecutive in the final sorted permutation,
their ranks form a contiguous interval, which we denote by $I_S$.

Because every partition is stable, the relative order of the elements of $S$
is preserved throughout the recursion. Therefore, if a recursive subproblem
corresponds to a canonical rank interval $J \subseteq I_S$, then all elements
of that subproblem belong to $S$ and already occur in increasing order. The
sortedness test succeeds at that node, and no further recursive calls are
made below it.

Consider the decomposition of $I_S$ into maximal canonical intervals
\[
B_1,B_2,\ldots,B_t,
\]
where each $B_r$ is contained in $I_S$, but its parent interval in the
partition tree is not contained in $I_S$. Let
\[
b_r = |B_r|.
\]
The intervals $B_1,\ldots,B_t$ are disjoint and cover $I_S$, so
\[
\sum_{r=1}^t b_r = m.
\]

Each canonical interval $B_r$ becomes a terminal recursive subproblem. 
An element of $B_r$ participates in at most
\[
\log_2 \frac{n}{b_r} + O(1)
\]
recursive levels before reaching $B_r$. At each active recursive node, the
sortedness test, median selection, and stable partitioning take linear time
in the size of the node. Charging this work uniformly to the elements of the
node gives $O(1)$ charge per element per level. Hence the total charge to the
elements of $B_r$ is
\[
O\!\left(
b_r\left(\log\frac{n}{b_r}+1\right)
\right).
\]

This quantity is summed over all the maximal canonical intervals inside
$I_S$. We use the standard property of dyadic decompositions that there are
at most two maximal canonical intervals of any given size. Therefore,
\[
\begin{aligned}
\sum_{r=1}^t b_r \log\frac{n}{b_r}
&=
\sum_{r=1}^t b_r
\left(
\log\frac{n}{m}+\log\frac{m}{b_r}
\right)\\
&=
m\log\frac{n}{m}
+
\sum_{r=1}^t b_r\log\frac{m}{b_r}.
\end{aligned}
\]
%%%%%%%%%%%%%%%%
Let
\[
2^p \le m < 2^{p+1},
\]
and, for each \(j\in\{0,1,\ldots,p\}\), let
\[
c_j
=
\left|
\left\{
r\in\{1,\ldots,t\}: b_r=2^{p-j}
\right\}
\right|.
\]
Since \(B_1,\ldots,B_t\) form the maximal canonical decomposition
of \(I_S\), there are at most two blocks of each size; hence
\(c_j\le 2\). Grouping the blocks according to their sizes gives
\[
\begin{aligned}
\sum_{r=1}^{t} b_r\log_2\frac{m}{b_r}
&=
\sum_{j=0}^{p}
c_j\,2^{p-j}
\log_2\frac{m}{2^{p-j}}.
\end{aligned}
\]
Since \(m<2^{p+1}\),
\[
\log_2\frac{m}{2^{p-j}}<j+1.
\]
Consequently,
\[
\begin{aligned}
\sum_{r=1}^{t} b_r\log_2\frac{m}{b_r}
&\le
\sum_{j=0}^{p}
2\cdot 2^{p-j}(j+1)\\
&=
2^{p+1}\sum_{j=0}^{p}\frac{j+1}{2^j}\\
&\le
2^{p+1}\sum_{j=0}^{\infty}\frac{j+1}{2^j}\\
&=
2^{p+3}
=
O(m),
\end{aligned}
\]
where the last equality follows from \(2^p\le m\).
%%%%%%%%%%%%%%%
Consequently, the total work charged to $S$ is
\[
O\!\left(
m\log\frac{n}{m}+m
\right)
=
O\!\left(
m\log\left(\frac{n}{m}+1\right)
\right).
\]

Finally, the maximal $\alpha$-sorted subsequences form a disjoint partition
of the input. Summing the above bound over all such subsequences yields
\[
O\!\left(
\sum_{i=1}^k
n_i \log\left(\frac{n}{n_i}+1\right)
\right).
\]

The algorithm never computes the maximal $\alpha$-sorted subsequences or
their lengths, so no prior knowledge of the values $n_i$ is required.

If $n$ is not a power of two, balanced median partitions produce canonical
intervals whose sizes differ by at most one at each level. The same argument
then applies with only constant-factor changes.
\end{proof}

\ignore{
%\begin{lemma}
%Consider a maximal $\alpha$-sorted sub-sequence ${\cal S}$ of length
%$n_j$ - where $2^p \leq n_j < 2^{p+1}$. Then, there will be a consecutive 
%pair of elements $( M^{\log n - (p-1)}_s, M^{\log n - (p-1)}_{s+1}) \in {\cal S}$ for 
%some $s$. Moreover there is no such pair in level $< \log n - p$. 
%\end{lemma}
This follows from the fact that the elements of ${\cal S}$ are consecutive in the sorted
output. So any interval greater than $2^p$ must contain a pair of elements that are
separated by $2^{p-1}$. Figure \ref{consec} illustrates this property. 
There cannot be consecutive such pairs in level $\log n - (p +1)$
since $| {\cal S} | < 2^{p+1}$; however, there may be consecutive pairs from level
$\log n -p$, that will not be required in the analysis. 
\begin{figure}[h]
\includegraphics[width=5in]{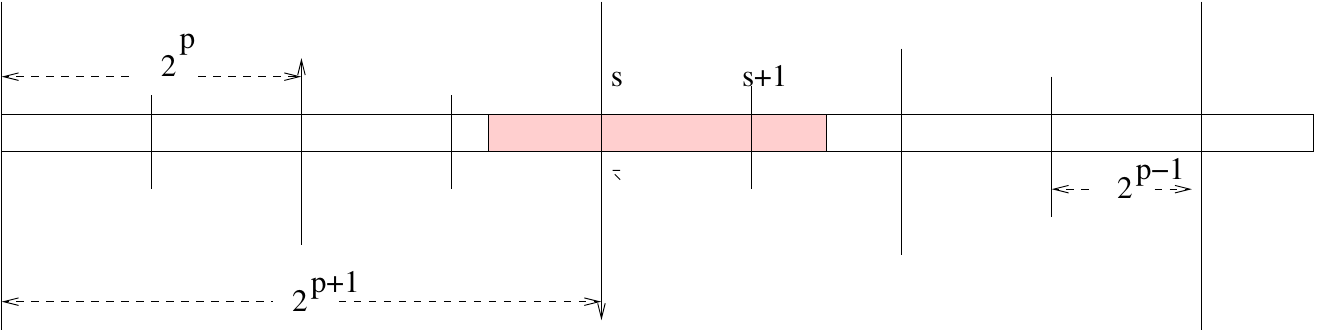}  
\caption{The figure shows overlays of partitions of sizes 
$2^{p+1} , 2^p , 2^{p-1}$. The shaded interval ${\cal S}$ has length $L$ where
$2^{p+1} > L > 2^p$ and there are two consecutive 
elements $[s, s+1]$ 
of partition $2^{p-1}$
that must fall within ${\cal S}$.    
} 
\label{consec}
\end{figure} 
Intuitively, we are trying to identify the largest subset of ${\cal S}$ that 
does not
incur further recursive calls (not proceed past Step 2). The above 
observation and the
basic property of partition sort, every element moves into the appropriate $2^{\log n -p}$ 
sized interval after $p$ recursive calls, will lead us to the desired result.
 
\begin{proof} (of Theorem \ref{optsort}) 
By the time, the recursive calls are made on inputs of size $2^{p}$, where
$2^p \leq n_j < 2^{p+1}$, at least $\frac{n_j}{2}$ elements 
will be contained in a subproblem. Note that, in the subsequent recursive call, the 
(contiguous) interval of size $2^{p-1}$ will be sorted and no further recursive call is
required. By charging each of the $n_j$ elements for the comparisons done for the
partitioning step over $\log n - \log ( n_j /2) = \log ( \frac{2 n}{ n_j })$ levels,
the $2^{p-1}$ elements will incur at most $\frac{n_j}{2}\log ( \frac{2 n}{ n_j })$
cost. The remaining $n_j - 2^{p-1}$ elements will be contained in at most two $2^{p-1}$
blocks (adjacent to the sorted block). Suppose there are $r$ elements in the right
adjacent block, then $r = \sum_{i=1}^{p-1} a_i \cdot 2^{p-1 -i}$ for $a_i = 0/1$ 
corresponding to the binary representation of $r$.
Then, the previous charging argument yields 
\begin{align*}
\sum_{i=1}^{p-1} a_i \cdot 2^{p-1 -i} \log ( \frac{n}{2^{p-1 -i}}) & =  
  \sum_{i=1}^{p-1} a_i \cdot 2^{p-1 -i}\log \frac{n}{n_j}  + 
\sum_{i=1}^{p-1} a_i \cdot 2^{p-1 -i} \cdot i \\
   & = O( n_j \log ( n/n_j )) + O( n_j)
\end{align*}
using $2^p \leq n_j < 2^{p+1}$.
This can be repeated for the other adjacent block
and using similar calculations for the other $n_i$s, 
the total number of comparisons is 
\[ \sum_{i}^{k} O( n_i \log ( \frac{n}{n_i} + 1 )) \].
\end{proof}
} 
\begin{remark} 
(i) For fewer sub-sequences, for example, if we can club two sub-sequences into a single
longer subsequence, the running time improves because of the sub-additive property of $H()$. 
This
shows the relevance of $\alpha$-sorted sub-sequences to the degree of sortedness. Moreover,
the maximal property will be crucial for the stronger lower-bound arguments.\\ 
(ii) This bound is very similar to multisorting \cite{MS:76} that is now 
generalized 
to a larger class of inputs. The elements that 
have identical values is a maximal $n_i$-sorted subsequence, although
it need not be a stable multisorting. \\  
(iii) 
As an alternative to exact median, we can use a randomized
median finding algorithm like \cite{FR75} that succeeds with high probability.
In a later section, we analyze the performance of a randomly chosen pivot 
similar to traditional Quicksort.
\end{remark}
\section{A lower bound for $(n_1 , n_2 \ldots n_k)$-sorted input}
 
To prove a matching lower bound, we will find a bound on the number of permutations of
$\{ 1, 2 \ldots n\}$ that have maximal sorted sub-sequences of lengths 
denoted by a $k$-tuple $( n_1 , n_2 , \ldots 
n_k )$.
We will consider any permutation of this $k$-tuple as the same {\it sorted type} 
and will consider all the distinct permutations of $\{ 1, 2 \ldots n\}$ 
that belong to the same sorted-type.
For example if the sorted sequence is $(1,2,3,4,5)$, then $(4 , 1, 2, 5, 3)$ and 
$(1 , 4, 2, 5, 3 )$ are distinct even though they have the same sorted sub-sequences of
lengths $(3, 2)$. Moreover, $(3, 1, 4, 2, 5)$ is also a distinct permutation of the same
{\it sorted-type}. The crux of the lower bound is an information
theoretic argument based on counting the number of valid permutations of the
same {\it sorted-type}. 

We will consider the following two scenarios for our lower bound 
\begin{itemize}
\item 
$k=\lfloor\gamma n\rfloor$ for any fixed constant $0<\gamma<1$\\
\item $n_i \geq 2 , \ \ \sum_i n_i = n$, implying $k \leq n/2$
\end{itemize}
The above two cases do not cover all possible inputs 
and consequently, our algorithm in 
its current form is not optimal for all inputs. 
For the first case, our lower bound is dependent on 
$k$ and is weaker than the ideal  lower bound for all sorted types with $k$ 
components. The crux of the problem is the case of $n_i = 1$ for a 
large number of $i$'s that our algorithm is unable to adapt to.   
%We will first tackle $k \geq n/2$.
\subsection{A lower bound based on $k$}
%%%%%%%%%%%%%%%%%%%
\begin{lemma}
Let $N_{n,k}$ denote the number of permutations of
$\{1,\ldots,n\}$ having exactly $k$ maximal
$\alpha$-sorted subsequences. Then
\[
N_{n,k}
\ge
\left(\min\{k-1,n-k\}!\right)^2.
\]
Consequently, for every fixed constant $0<\gamma<1$ and
$k=\lfloor\gamma n\rfloor$,
\[
\log N_{n,k}=\Omega(n\log n).
\]
\label{long-seq}
\end{lemma}

\begin{proof}
For a permutation $\pi$, let
\[
q_r=\pi^{-1}(r), \qquad r=1,\ldots,n.
\]
A consecutive set of ranks
\[
a,a+1,\ldots,b
\]
forms an $\alpha$-sorted subsequence of $\pi$ if and only if
\[
q_a<q_{a+1}<\cdots<q_b.
\]
It follows that the maximal $\alpha$-sorted subsequences of $\pi$
are in one-to-one correspondence with the maximal increasing runs
of the inverse permutation $q=\pi^{-1}$. Hence
\[
K(\pi)=1+\operatorname{des}(\pi^{-1}).
\]
where $\operatorname{des}(q)$ denotes the number of descents of $q$.
Since the map $\pi\mapsto\pi^{-1}$ is a bijection, the number of
permutations with exactly $k$ maximal $\alpha$-sorted subsequences is
the Eulerian number (see \cite{GKP:94})
\[
\left\langle {n\atop k-1}\right\rangle .
\]

For $k=\lfloor\gamma n\rfloor$, where $0<\gamma<1$ is fixed,
standard estimates for Eulerian numbers give
\[
\left\langle {n\atop k-1}\right\rangle
=
\exp\!\left(n\log n-O_\gamma(n)\right).
\]
%%%%%%%%
%%%%%%%%For a permutation $\pi$, let
%%%%%%%%\[
%%%%%%%%q_r=\pi^{-1}(r),\qquad r=1,\ldots,n.
%%%%%%%%\]
%%%%%%%%The maximal $\alpha$-sorted subsequences of $\pi$ correspond to
%%%%%%%%the maximal increasing runs of $q$. Hence $\pi$ has exactly $k$
%%%%%%%%maximal $\alpha$-sorted subsequences if and only if $q$ has exactly
%%%%%%%%$d=k-1$ descents.
We also provide an elementary direct proof.
First suppose that $d\le (n-1)/2$, and let $r=n-2d$. Define
\[
L=\{r+1,\ldots,r+d\},
\qquad
H=\{r+d+1,\ldots,n\}.
\]
Choose arbitrary permutations
\[
(\ell_1,\ldots,\ell_d)
\quad\text{of }L
\]
and
\[
(h_1,\ldots,h_d)
\quad\text{of }H,
\]
and form
\[
q=
1,2,\ldots,r,
h_1,\ell_1,h_2,\ell_2,\ldots,h_d,\ell_d.
\]
Since every element of $H$ is larger than every element of $L$,
the only descents of $q$ are
\[
h_i>\ell_i,\qquad i=1,\ldots,d.
\]
Thus $q$ has exactly $d$ descents. The two orderings can be chosen
independently, giving $(d!)^2$ such permutations.

Now suppose that $d>(n-1)/2$. Set
\[
d'=n-1-d=n-k.
\]
Then $d'<(n-1)/2$. By the previous construction, there are at least
$(d'!)^2$ permutations with exactly $d'$ descents. Replacing every
entry $x$ by $n+1-x$ interchanges ascents and descents, and therefore
maps each such permutation to one with exactly
\[
n-1-d'=d
\]
descents\footnote{Note that we have no control on the $n_i$'s}. Hence
\[
N_{n,k}\ge
\left(\min\{k-1,n-k\}!\right)^2.
\]

For $k=\lfloor\gamma n\rfloor$, where $0<\gamma<1$ is fixed,
$\min\{k-1,n-k\}=\Theta(n)$. Stirling's approximation therefore
gives
\[
\log N_{n,k}
\ge
2\log\left(\min\{k-1,n-k\}!\right)
=
\Omega(n\log n).
\]
\end{proof}
\begin{remark}
(i) Note that for sorted type $( n_1 , n_2 \ldots n_k )$, 
there are $\Omega ( k) $ subsequences with
$O(1)$ elements. The entropy bound in Theorem \ref{optsort} for 
$k=\lfloor\gamma n\rfloor$, where $0<\gamma<1$ 
yields $O ( n\log n  ) $ and so it matches this lower bound.\\
(ii) The above lower bound is not tight for $k \in n - o(n)$. Indeed, 
for the sorted-type $(1,1, \cdots 1)$,
the reverse-sorted permutation is the unique permutation that 
satisfies this. Since our algorithm takes $\theta ( n \log n)$ 
steps for this input, it is clearly not optimal for this sorted type. 
\end{remark}
%%%%%%%%%%%%%%%%%%%%

\subsection{\bf The case of $n_i \geq 2$} 

From the sorted permutation, we can pick any of the ${ n \choose n_1 }$ sub-sequences, and
subsequently ${ (n - n_1 ) \choose n_2 }$ sub-sequences and so on. This results in the   
multinomial expression ${\cal N} = \frac{n!}{ n_1 ! \cdot n_2 ! \ldots n_k ! }$. 
However, it may end up counting non-maximal sorted-types. To eliminate that,
we can use the property that if $n_i , n_j$ are two sorted sub-sequences that can make a 
longer subsequence in the permutation, they must be in order, viz., all elements of
subsequence $n_j$ must follow all the $n_i$ elements in the permutation. This can be adjusted
by swapping the last and the first element without changing the relative ordering of the 
$n_i$ and $n_j$ elements. This may be required between other pairs also, i.e., upto
$k -1$ exchanges to transform to a valid maximal {\it sorted-type}. 
\begin{figure}[h] 
\begin{tabular}{|c|c|c|c|c|c|c|c|c|c|c|c|c|c|c|c|}
\hline
4 & 6 & 8& 23 & 25 & 33 & 34 & 39 & 45 & 55 &
56 & 62 & 68 & 72 & 84 & 85 \\
\hline
6 & \cellcolor{pink} 23 & 8 & \cellcolor{pink}  25 & 84 & \cellcolor{pink} 33 &\cellcolor{cyan}
 62 &
\cellcolor{pink} 34 & \cellcolor{pink} 39
 & 85 &\cellcolor{cyan} 68 & \cellcolor{yellow} 45 & \cellcolor{green} 
72 & \cellcolor{yellow}  55& 4 & \cellcolor{yellow} 56 \\
\hline
6 & \cellcolor{pink} 23 & 8 & \cellcolor{pink}  25 & 84 & \cellcolor{pink} 33 & \cellcolor{cyan}
62 &
\cellcolor{pink} 34 & \cellcolor{yellow} 45
 & 85 &\cellcolor{green} 72 & \cellcolor{pink} 39 & \cellcolor{cyan}
68  &\cellcolor{yellow}  55& 4 & \cellcolor{yellow} 56 \\
\hline
\end{tabular}
%\\
\caption{
We are interested in a maximal 2-5-2-2-1-3-1 sequence that could correspond to
(6,8) (23,25,33,34,39 ) (84,85) (62,68)(72)(45,55,56)(4)
sub-sequences, but in row (ii), 45 follows 39 and 72 follows 68, thus making it into a
maximal 2-8-2-3-1 sequence. In row (iii) we have swapped 45 with 39 as well as 72 with 68
so as to transform it into a legitimate 2-5-2-2-1-3-1 maximal sequence.}
\label{mapping}
\end{figure}

\ignore{
Let us consider a
bipartite graph with a set of vertices $A$ representing ${\cal N} $ 
non-maximal sorted-types and a set $B$ representing $\nu$ valid 
maximal sorted-types. We will use the edges to define a mapping between 
$A$ and $B$ in the following manner. 

Consider any non-maximal decomposition $D \in A$: from Lemma \ref{disj}, 
there is a unique maximal
$\alpha$-sorted subsequence decomposition ${\cal M}$ corresponding to $D$. If
$D$ is not maximal, then there must be a maximal subsequence of ${\cal M}$
that is present in $D$ as multiple subsequences in order. We
will locally swap elements from consecutive subsequences ${\cal M}$ (see Figure
\ref{mapping} for an illustration). We can continue doing this to the
transformed permutation $D'$ and the corresponding ${\cal M'}$. Since the
maximal subsequences of ${\cal M}$ are disjoint, this does not affect other
sequences and we can do these transformations repeatedly.  

By swapping some appropriate subset of the $k-1$ pairs a non-maximal 
sorted-type is mapped
to a valid maximal-sorted type. Each edge can be labelled using $d_m$ swap 
locations 
where the number of possibilities $ d_m \leq 
{n \choose 2k}$. 
\begin{lemma}
If $S_1, S_2$ be two non-maximal $\alpha$-sorted subsequences, then they
 cannot map to the same maximal $\alpha$-sorted subsequence using identical
swap transformations. 
\end{lemma}
Indeed, each transformation has a unique inverse, so $S_1 \neq S_2$ cannot
map to the same subsequence.
It follows that any valid maximal sorted-type can 
have a degree bounded by $d_m$, for distinct non-maximal permutations. 
In the bipartite graph we are looking at a degree bound of $\leq d_m$ on 
set $B$ and at least 1 on set $A$ giving us $|A| \leq \#Edges \leq d_m \cdot 
|B|$.
}%ignore

\begin{lemma}
Let $n_1+\cdots+n_k=n$, where $n_i\ge2$ for every $i$, and let
$\mathcal A$ be the set of permutations admitting the prescribed
$\alpha$-sorted decomposition $(n_1,\ldots,n_k)$. Let $\mathcal B$
be the subset for which all prescribed subsequences are maximal.
Then
\[
|\mathcal B|
\ge
\frac{1}{2^{k-1}}
\frac{n!}{n_1!\cdots n_k!}.
\]
\end{lemma}

\begin{proof}
Let
\[
s_i=\sum_{j=1}^i n_j,
\qquad 1\le i<k,
\]
so that the $i$-th prescribed boundary lies between the consecutive
ranks $s_i$ and $s_i+1$.

For $\pi\in\mathcal A$, define its signature by
\[
\Sigma(\pi)
=
\left\{
i:
\operatorname{pos}_{\pi}(s_i)
<
\operatorname{pos}_{\pi}(s_i+1)
\right\}.
\]
These are exactly the boundaries at which two adjacent prescribed
blocks can merge.

For every $i\in\Sigma(\pi)$, swap the input positions of the
consecutive-rank elements $s_i$ and $s_i+1$. Since
\[
s_{i+1}-s_i=n_{i+1}\ge2,
\]
the pairs associated with distinct boundaries are disjoint, and
the swaps can be performed in parallel.

Each swap preserves the increasing order within both prescribed
blocks: the last element of the left block moves later, and the
first element of the right block moves earlier. At the same time,
the swap reverses the relative order of the consecutive-rank
elements at that boundary. Thus the two blocks can no longer merge.
Boundaries not in $\Sigma(\pi)$ are already non-mergeable and are
unaffected by swaps at other boundaries. Hence the resulting
permutation $F(\pi)$ belongs to $\mathcal B$.

Construct a bipartite graph with left vertex set $\mathcal A$,
right vertex set $\mathcal B$, and an edge
\[
\pi\longrightarrow F(\pi)
\]
labelled by $\Sigma(\pi)$. Every left vertex has degree one.

Fix $\tau\in\mathcal B$ and a signature
$\Sigma\subseteq\{1,\ldots,k-1\}$. Since each swap is an
involution, $\tau$ and $\Sigma$ determine at most one possible
preimage: apply the swaps indexed by $\Sigma$ to $\tau$. Therefore
every right vertex has degree at most
\[
2^{k-1}.
\]
Consequently,
\[
|\mathcal A|
\le
2^{k-1}|\mathcal B|.
\]
Finally,
\[
|\mathcal A|
=
\frac{n!}{n_1!\cdots n_k!},
\]
which proves the result.
\end{proof}
Appendix A illustrates this construction for a concrete example.
We can now lower 
bound the number of permutations of the maximal sorted-type for $k \leq (n/4)$
as follows 
\begin{equation}
  \log \left( | \mathcal B | \right) \geq \log \left( \frac{| \mathcal A |}{2^{k-1}} \right)  =
 n \cdot H \left( \frac{n_1}{n}, \frac{n_2}{n} \ldots \frac{n_k}{n}\right) 
-(k-1) \in \Omega \left( n \cdot H \left( \frac{n_1}{n}, \frac{n_2}{n} \dots
\frac{n_k}{n}\right) \right )
\label{lbndeqn}
\end{equation}
 
\ignore{
By taking the logarithm of the above 
expression, $\log | \mathcal B |$ can be
lower-bounded by $\Omega ( n \cdot H ( \frac{n_1}{n}, \frac{n_2}{n} \dots 
\frac{n_k}{n})$ for $H ( \frac{n_1}{n}, \frac{n_2}{n} \dots
\frac{n_k}{n}) > c$ for some appropriate constant $c > 1$. 
%See appendix for a detailed calculations.
		For $k = o(n)$, we can use the Stirling's approximation $ {(\frac{n}{k})}^{k} \leq 
		{n \choose k} \leq {(\frac{n e}{k})}^{k}$ and for $k = \Omega (n)$, one
		can show $\log ( {n \choose k}) \leq (1 + o(1)) n$ \cite{Das}. 

		We will assume wlog that $H ( \frac{n_1}{n}, \frac{n_2}{n} \dots \frac{n_k}{n}) > c'$. This
		will be good for any lower-bound for $\log 
		| \mathcal B | = \Omega ( \max\{ n H() , c'n \}) = \Omega (n)$ that
		suffices for our purpose. 

		Note that for $k = \Omega (n)$, the expression 
$\log \left( { n \choose 2k} \right)$ can be bounded by $1.01 n$.
		Therefore, $H ( \frac{n_1}{n}, \frac{n_2}{n} \dots \frac{n_k}{n}) > c$, 
		and the required 
		bound trivially holds.

		For $k = o(n)$, taking the logarithm of both sides of the Equation \ref{lbndeqn},
		we obtain,

		\[ | \mathcal B | \geq \frac{| \mathcal A | }{{n \choose 2k}} =
		\frac{n! }{ n_1 ! \cdot n_2 ! \ldots n_k ! } \cdot \frac{ (n-2k)! \cdot (2k)!}{ n!} \]
		and applying Stirling's approximation,
		\[ \log | \mathcal B | \geq  \sum_{i=1}^{k} ( n_i \log \frac{n}{n_i})  -2k \log (\frac{ne}{2k}) \] 

		Since $2k \log (\frac{ne}{2k})$ is bounded by $O(n)$, the required
		lower-bound for $\log \nu$ follows. 
		%We will now show termwise on the R.H.S. that 
		%\[ n_i \log \frac{n}{n_i}  -2 \log \frac{ne}{2k} \geq \Omega ( n_i \log \frac{n}{n_i} ) \]
		%For concreteness, let us choose $ n_i /2 \cdot \log \frac{n}{n_i}$, so the above is equivalent
		%to showing that for all $i , \ \ n_i /4 \cdot  \log \frac{n}{n_i} \geq \log \frac{n}{k} $.
		%
		%Note that, we can also assume that $k \geq c'$ for some constant $c'$, otherwise, 
		%$n H() = O(n)$. For concreteness, say $k \geq 16$. 
		%
		%Case :  $ 4 \leq n_i < k$, we have $ n_i /4 \cdot \log \frac{n}{n_i} \geq \log \frac{n}{k}$  
		%\\
%Case : $n_i \geq k = 8$, we get $n_i /4 \log \frac{n}{n_i} - \log \frac{n}{k}$  
}%ignore
Therefore, we have a lower bound that is applicable
to any fixed sorted type in line with the bound of Theorem \ref{optsort}.
\begin{theorem}
An input consisting of $( n_1 , n_2 \ldots n_k )$-sorted sub-sequences, where
$ n_i \geq 2 , \sum_i n_i = n $,
requires $\Omega ( n \cdot H ( \frac{n_1}{n},
\frac{n_2}{n} \dots \frac{n_k}{n}) + n)$ comparisons. 
\label{lbnd}
\end{theorem}
\begin{remark}
(i) The restriction $n_i \geq 2$ stems from the difficulty of dealing with a 
large number
of size 1 maximal subsequences. These elements form a decreasing subsequence 
that limits the number of permuations. As we have observed before, 
the sequence is reverse sorted for the case of $n_i =1$ for all $i$. 
\\
%(ii) The lower-bound expression for multisorting in \cite{MS:76} has a 
%negative term
%$(n-k)\log\log k$ since it does not guarantee stability.\\
(ii) There is a gap between the upper and lower bound for $k \geq 
n - o(n)$
that can be addressed to some extent by a modification of the algorithm. 
\end{remark}
Overall, the gap is
 related to some interesting open problems involving nearly sorted 
sequences that we briefly discuss here. 
In general, a large value
of $k$ implies a high degree of reverse sorted sequence that the partition sort
is unable to leverage in its current form. 
This can be addressed by replacing the Step 4 of Algorithm Stable-Partition-sort 
with a more sophisticated check. Instead of just checking if it is fully sorted, 
we can check for {\it nearly} reverse sortedness property 
using the algorithm
of Levcopoulos and Petersson \cite{LP:94}. Their algorithm called {\it slabsort},
sorts a sequence consisting of $m$ increasing or decreasing sorted subsequences
in $O(n\log m)$ comparisons. For $\log m = o(\log nn)$, this 
would be $o(n\log n)$, for 
example, if $k = O(polylog n)$.\\
We also observe that it is related to identifying the longest decreasing
subsequence problem. If the reverse sorted sequence ${\cal L}$
 has length $n - o(n)$,
and we can identify this in $o( n\log n)$ comparisons, we can complete 
sorting in $o(n \log n)$ time. This can be achieved by 
inserting the $o(n)$ out of sequence elements in ${\cal L}$.
The known lower bound for this problem \cite{Fred:75}  is 
$\Omega ( n+ (n - 2L )\cdot \log L)$ 
where $L$ is the length of the longest monotonically increasing 
subsequence. 
The $\Omega (n\log n)$ lower bound for the longest increasing subsequence problem
is obtained by using $L = \frac{n}{\log n}$ which is covered by the previous
Lemma.\\
For $L = n/2 - o(n)$, the lower bound is $\Omega (n)$. 
To the best of our knowledge, this remains a very challenging open problem to
establish a tight bound.  \\
\section{Quickersort - randomized stable adaptive Quicksort }

The algorithm given in Figure \ref{partitionsort} uses a linear time 
median finding sub-routine that may slow down the implementation by a constant
factor. It would be more practical to use a randomized median finding or 
choose a random pivot like randomized Quicksort. The rest of the algorithm
is identical, but it cannot be guaranteed that the partitions will 
be shrinking by a factor of 2. 
\begin{figure}[h]
\begin{procedure}[H]
 \nl {\bf Input:} An unordered array $A$ of $n$ numbers $x_1 , x_2 \ldots x_n$  \;
 \nl \If {$A$ is not sorted (verify by comparing adjacent elements)}{
\nl Choose a uniformly chosen random element $M$ as pivot, and partition $A$    around $M$ using a stable procedure \;
 Quickersort ( $A_< $) /* $A_<$ : elements smaller than $M$ */ \;
 Quickersort ( $A_> $) /* $A_<$ : elements greater than $M$ */ \;
}
\nl {\bf Output} $A$
\caption{Quickersort($A$) }
\end{procedure}
\caption{ Stable version of Partition Sort}
\label{quickersort}
\end{figure}

There are several analyses of randomized Quicksort in
the literature \cite{KR93,Seidel92,Mc96}. In the remaining section, we will
assume that the random pivots have been generated by an initial random 
permutation, viz., the $p$-th pivot is the $p$-th element of the random 
permutation and splits the interval that it belongs to. 

We will base it on a simple 
dart throwing game \cite{Mu93} that
is defined as follows - \\
Imagine that an observer is standing at the origin
of a real line and throwing $m$ darts at random locations $\{ 1, 2, \ldots 
m \}$
in the positive
direction. When a dart is thrown, only the closest dart is visible to the
observer. After all the $m$ darts have been thrown, we are interested in the
number of darts that were visible to the observer during the course of the
experiment. For example if a dart landed in location $i$ before location $j$
and $i < j$, then only $i$ would be seen. However if $j$ was thrown before $i$,
then both darts would be observed. Let $X$ be the random variable denoting the 
number of darts visible to the observer that is dependent on the order they
are thrown.
\begin{lemma}
\[ \E [X] = H_m = \sum_{i=1}^m \frac{1}{i} \] 
\label{dartgame}
\end{lemma}  
This follows from the observation that for the dart in position $i$ to be 
visible, none of the darts in positions $1 \ldots (i-1)$ can be thrown 
before $i$ and the probability of that is $\frac{1}{i}$.

We can approximate $H_m $ by $\log m$. The process of randomized Quicksort 
can be analysed by observing that any fixed element $E$ belongs to an interval
$J_i = [\ell_i , r_i ]$ after $i$ pivots have been chosen and by convention
$\ell_0 = - \infty , r_0 = + \infty$. When the next pivot $M$ is chosen, 
$E$ will be compared to $M$, iff $M \in J_i$ (assuming $M \neq E$. If $M, E$
are compared, then $J_{i+1} \neq J_i$ and one of the end-points will change. 
A dart can be thought of
as a random pivot and only those pivots are compared with an element that is
visible to it.
So the number of times $E$ is compared to a pivot can be mapped to the
dart-throwing game where the number of darts visible to its right (left)  
is equal to the number of times the right end-points (respectively the left
end-points) have changed. 
We can conclude that for rank $i$ element, the number of times it changes its 
partition during the execution 
(induced by the pivots) is $\log i + \log (m-1 - i)$. This is 
equal to the number of comparisons involving $i$-th element. Therefore the
expected number of comparisons is bounded by $2 \log n$ for any element
that results in $O(n\log n)$ expected running time.

\begin{figure}[h]
\begin{center}
\includegraphics[width=5in]{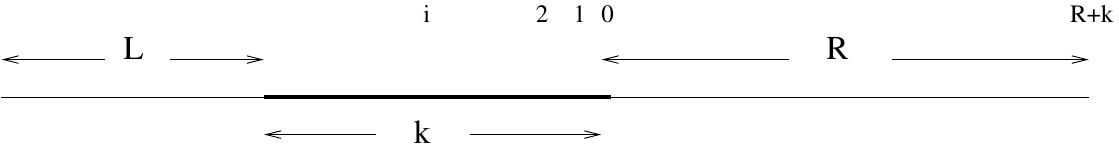} 
\caption{ The dart game analogy for the $k$-sorted subsequence 
framework where $n$ elements are in sorted order. 
The subsequece ${\cal S}$ are in consecutive $k$ locations and the 
number of elements to the left and right of 
${\cal S}$ are $L$ and $R$ respectively. The observer is at location 
$i$ of ${\cal S}$.} 
\label{dart}
\end{center}
\end{figure} 
In our set up, we would like to bound the expected number of comparisons 
involving an element $E$ that belongs to a sorted subsequence ${\cal S}$ 
of length $n_i$. 
The counting (for charging comparisons) of pivots will be categorized into 
left and right pivots for any given element $E$ and further subdivided into
the pivots within and outside of ${\cal S}$. 
Here $E$ will be compared to elements until it belongs to a partition
$\ell^* < E < r^*$ and $\ell^* , r^* \in {\cal S}$. Since all the 
elements in the 
interval $[ \ell^* , r^* ]$ are sorted, there will be no further comparisons.

Using Lemma \ref{dartgame}, we can bound the expected number of comparisons   
involving $E$ by $\sum_{j=i}^{i+R} \frac{1}{j} + 1$ where the additive 
1 represents the first time
a pivot is chosen from $[1, i ]$ (see Figure \ref{dart}). This is bounded
by 
 \[ \ln (R+i) - \ln i = \ln \left( \frac{R +i}{i} \right) = 
\ln \left( 1 + \frac{R}{i} \right) \]
the expected number of comparisons for element $i$ by pivots to its right.
We can obtain the expected number of comparisons for all elements by the sum
\begin{eqnarray*}
 \sum_{i=1}^{k} \ln (1 + \frac{R}{i}) & \leq & k + \sum_{i=1}^{k} \ln (\frac{R}{i})  \\
  & \leq & k \ln R - \ln( k !) \\
 & \leq & k + k\ln R - k\ln k + O(k) \text{ from Stirlings approximation } \\
 & \leq & 2k + k\ln \left( \frac{R}{k} \right) 
\end{eqnarray*}  

Similarly we can bound the comparisons from the left pivots by 
 $2k + k\ln (\frac {L}{k} )$. By adding the two terms, we obtain 
$4k + k\ln (\frac{R}{k} ) + k\ln (\frac {L}{k} )$, where $L+R + k = n$. This
is maximized where $R=L = \frac{n-k}{2}$ giving us a bound of
\[ 4k + 2k\ln ( \frac{n-k}{k}) - 2k\ln 2 \]
 For $k = o(n)$, this is $O(k) + 2k \ln \frac{n}{k} $ and for $k = \theta (n)$
this is $O(k)$. 

The above calculation counts the number of pivots outside the subsequence 
${\cal S}$, the elements $E$ will continue to be compared to a pivot until both
$\ell_i$ and $r_i$ belong to ${\cal S}$ and these may not happen simultaneously.
Even if $\ell_i \in {\cal S}$, there will be more pivots to the left of 
$E$ till
$r_j \in {\cal S}$ for $j > i$. So, we will now account for the 
pivots within ${\cal S}$. 
\begin{lemma}
Consider the $j$-th element of ${\cal S}$ from the right. The expected number
of pivots picked within ${\cal S}$ before picking a pivot 
numbered lower than $j$ is $\log \left( \frac{k}{j} \right)$. 

\end{lemma}

\begin{proof} Our calculation is conditioned on the pivots chosen from the 
$k$ elements in ${\cal S}$. Since the pivots are chosen uniformly at
random one can argue that for the
$j$-th element where $1 \leq j \leq k$, the probability that a pivot is
to the right is $\frac{j}{k}$. Therefore the number of pivots chosen from the
left before choosing from the right is a random variable $Y$ having 
a geometric distribution whose
expectation is $\frac{k}{j}$. From Lemma \ref{dartgame}, we know that only
$\log Y$ will be visible to the $j$-th element in expectation. 
This can be upper bounded using an application of Jensen's inequality 
to the $\log$ function that is a concave function, yielding a
bound of $\log \left( \frac{k}{j} \right)$.  
\end{proof}
Consequently, the total number of
comparisons for all the elements in ${\cal S}$ can be bounded by    
\begin{eqnarray*}
 \sum_{j=1}^{k} \log (\frac{k}{j} ) & = & k \log k - \log ( k! )   \\
  & \leq & k \ln k - ( k \log k - k \log e + O(\log k) 
\text{ from Stirlings approximation } \\
 & \leq & 2k
\end{eqnarray*}

By using $k = n_i$, for each $\alpha$-sorted subsequence of length $n_i$,
the total expected number of comparisons due to pivots within ${\cal S}$ 
and outside ${\cal S}$ can be summarized as follows. 
\begin{lemma}
The expected number of comparisons involving elements in
any $n_i$-sorted subsequence can be 
bounded by $O( n_i + n_i \cdot \log \frac{n}{n_i} )$.
\end{lemma}

By summing over all the $n_i$-sorted subsequences, 
we can state the analogue of Theorem \ref{optsort} by 
replacing the explicit median finding
step for partitioning in Step 3 of Algorithm Stable-Partition-Sort by a
randomly chosen pivot. We will refer to this version of the algorithm 
as {\it Quickersort} to distinguish it from the traditional Quicksort.
\begin{theorem}
An input consisting of $( n_1 , n_2 \ldots n_k )$-sorted sub-sequences
can be sorted by Quickersort using an expected 
$O( \sum_{i=1}^{k} n_i \log ( \frac{n}{n_i} + 1 ) ) = O( n \cdot H ( \frac{n_1}{n} , 
\frac{n_2}{n} \dots \frac{n_k}{n}) + n)$ comparisons where $\sum_i n_i = n$.
\label{randoptsort}
\end{theorem}

\section{Concluding remarks and universal adaptivity }

The traditional running time analysis is measured as a function 
of only the input size $n$ which addresses the worst case behavior but
does not capture the running time of any specific instance 
of size $n$.
The term {\it adaptive algorithms} is often used to refer to algorithms that
exploit additional information for instances that can lead to faster 
(than worst case) behavior. 
In particular, we are interested in a
class of algorithms that is dependent on multiple parameters that
are not known a priori. Yet, we would like to achieve
performance comparable to
algorithms that are designed based on the knowledge of these parameters. 

\ignore{
For example, consider the fundamental geometric problem of 
computing the convex hull of $n$ points. In the planar case, the output size 
can vary 
between $O(1)$ to $n$ and it took researchers \cite{KS86} 
some time to understand the optimal
complexity of convex hull in terms of $n$ and output size $h$ which was shown to be 
$\Theta (n \log h)$. 
Alternate analysis given by \cite{SenG99} and further refined 
subsequently in \cite{ABC17} led to 
running times of $O( \sum_i^h n_i \log (n/ n_i + 1))$
and $n \cdot H^* ( n_1 / n , n_2 /n \ldots n_h /n )$ respectively\footnote{This subtle difference 
is out of scope of the present paper to distinguish},
both of which are upper-bounded by $O( n \log h )$. 
The vector ${\cal D} = ( n_1 , n_2 
\ldots n_h )$ is not an input parameter but related to the combinatorial 
structure of the final output and not part of the input. 
It may be intriguing at first to  
the reader that both the $O(n \log n )$ and $O(n\log h)$ algorithms 
for convex-hull were shown to be
optimal. The finer characterization
in terms of output, and even more so in terms of ${\cal D}$, 
leads to a superior
and deeper understanding of the complexity of
the problem for different instances of the same input size. By noting that
$\sum_i^h n_i \log (n/ n_i +1) \leq n\log h \leq n\log n$ 
one can appreciate the increasing
refinement of the complexity of an instance using additional parameterization. 
}

We consider {\em oblivious algorithms} to be a sub-class of
adaptive algorithms that neither has prior knowledge of many parameters nor 
do they explicitly try to estimate them using strategies like
 doubling or squaring.
{\it Oblivious} algorithms do not rely on the history or any input distribution
but on a finer characterization of the problem related to (unknown) implicit 
parameters. 
For example, we can design an $O(n\log d)$ sorting algorithm for inputs where 
every element
is within distance $d$ using a heap of size $d$, but it will not be considered
oblivious since it requires the knowledge of $d$ or an estimation of $d$.
For the related geometric problem of
planar convex hull of $n$ points, the worst case optimal bound of $O(n\log n)$ 
was improved to $O(n\log h)$ \cite{KS86} where the output size 
can vary between $O(1)$ to $n$. 
Further refinements were obtained by \cite{SenG99} and 
\cite{ABC17} that resulted in the 
running time of $n \cdot H ( n_1 / n , n_2 /n \ldots n_h /n )$ where
$n_i$'s are related to the distribution of the $h$ output points. 

More specifically, the 
analysis of the running time can be expressed as a function of such parameters 
and the goal is to understand the exact (or asymptotic) dependence on them.  
In order to address such problems, we consider all
the $n !$ permutations of an instance $|I| = n$. Consider an algorithm 
$A_{\sigma}$ that verifies if a certain permutation $\sigma$ is sorted, 
or else, it invokes a more general sorting algorithm.
This will be superior to any normal algorithm that has to 
deal with an arbitrary permutation and cannot match the performance for 
every instance. An algorithm $A$ is known as {\it instance optimal} if 
it satisfies
\[ \forall I , {|I| = n}, T_A (I) \leq \beta \min_{ A^* \in {\cal A}}
T_{A^*} ( I) \ \  \beta \mbox{ represents suboptimality factor
} \]
where $A, A^* \in {\cal A}$, the family of all correct algorithms for sorting.
If we set $\beta = O(1)$, instance optimal sorting algorithm will be 
required to have $O(n)$
time based on the family $A_{\sigma}$. This contradicts the worst case 
lower bound and therefore a more realistic 
framework can be defined as follows.
 
Let ${\cal C}(n)$ denote a classification of the $n!$ permutations
classified in terms 
of their measure of sortedness as per some predefined parameters. A given
input $I , |I| = n$, belongs to a specific class $\Pi'(n)$ where 
$\Pi' (n) \in {\cal C}(n)$ that may not be known to our algorithm, however, it 
is expected to be competitive with an algorithm that has this information.
In this context a universally
optimal algorithm $A$ is defined as  
\[ \forall I ,  |I| = n , T_A (I) \leq \beta  \min_{ A^* \in 
{\cal A}( \Pi' (n))}
\max_{ I \in \Pi' (n) } T_{A^* } ( I) \]
where ${\cal A}(\Pi' (n))$ denotes the class of algorithms customized for
$\Pi' (n)$. Although the left and right sides are defined for the same input, 
note 
that $\Pi' (n)$ is known to the right side, but not the left side, which makes it 
oblivious. Our choice of $\alpha$-sorted subsequence defines such a family of
permutations. Other measures of pre-sortedness can also be cast in the same
framework that compares the effectiveness of an algorithm w.r.t. 
subsets of instances rather than over the entire space of instances.  
%For instance,
%there is a fundamental difference between 
%\cite{Chan1995} that estimates the number of extreme points
%in contrast with \cite{KS86,BS97,CSY95} that converge in an oblivious manner.
%While it often works for the case of
%one parameter, it is not clear how to estimate multiple parameters. 
%Estimating parameters using sub-sampling has been used for sorting algorithms 
%\cite{CR93}. 
%More formally, we can define a hierarchy of parametrization as follows. Let 
%$N, N'$ represent a set of parameters (explicit or implicit) relevant to a
%specific problem ${\cal P}$ such that $N \subset N'$. Let $T_A ( N )$ and 
%$T_B (N) $ represent the running times of algorithms $A$ and $B$ for the same 
%instance, that are expressed as a function of $N , N'$ respectively.
%If $\max_{N' } T_B ( N' ) \in O ( T_A ( N ))$ for all instances and 
%$T_B ( N' ) \in o( T_A (N )$ for a family of instances,
%then we consider $N'$ to be a more {\it refined complexity} measure for the 
%problem ${\cal P}$.  

This framework can be easily extended to the recent notion of {\it universal 
optimality} in the context of Dijkstra's algorithm \cite{universal24} that
adapts to any graph topology and can match the performance of an algorithm
that is customized for any specific topology. This measure of optimality
is more useful than worst-case optimality as the algorithm can
adapt to specific classes of inputs and guarantee better performance. 
We feel that the natural candidates for universal optimal algorithms would
be the oblivious algorithms that do not require any estimation and can adapt
to even implicit parameters. Our analysis in this paper shows that 
randomized Quicksort has these properties.
\ignore{
Some interesting results have been recently discovered related to 
Dijkstra's Single source
shortest path algorithm. It was shown (in \cite{universal24}) 
using a more carefully designed heap data-structure, that the algorithm was 
optimal in a
very strong sense. The authors termed it as {\it universal optimality}
\cite{universal24}. There is a natural extension of the previous
framework that can capture universal optimality. 
 
For a problem $P$, let $A, A^* \in {\cal A}$ (set of
all correct algorithms ) and the input $I$ can be partitioned 
into two sets of parameters, $I' , I''$ where an algorithm can be specifically 
tailored for $I''$. We define a class of algorithms to be 
competitive with such a tailored algorithm as follows.  
\[ \forall I, {| I| = n }, T_A (I) \leq \beta \min_{A^* \in {\cal A}(I'')}
\max_{ I' } T_{A^* } ( I ) \mbox{ where } I = I' \cup I'' \]
Then $A$ is referred to as {\it universal optimal} within a factor $\beta $
and ${\cal A}( I'')$ represents the class of algorithms 
that has been customized for a specific $I''$. Under this framework, the
universally optimal algorithm 
in \cite{universal24}, has $I''$ capturing the graph topology and $I'$ 
corresponding to the edge weights. In general, $I$ can also capture implicit
parameters that could help in the analysis.
}
\ignore{This property is
a relaxation of the notion of {\it instance optimality} \cite{ABC17} where $I'$ is
empty and therefore it is like a fixed input for the customized algorithm.
These pertain
to a stronger notion of optimality than the traditional worst-case optimality where 
$I = I'$

Broadly speaking, such algorithms match the best possible algorithms for {\emph
 every}
instance of a problem within a constant (or a small multiplicative factor $\beta $). 
This shifts the focus of worst-case efficient 
algorithms 
to other instances for which we can attain better efficiency. 
The performance of such algorithms would be dependent on additional 
attributes of the specific problem instead of the size only. 
%The field of parametrized complexity
%quantifies the dependence of an optimization parameter on the running time but it lacks 
%comparable lower bounds. Hence it is not a good fit for our framework of universal optimality.

In our previous 
definition of universal optimality, we had ignored the dependence of the
algorithms on the ordering of input parameters in $I$, since we expressed
$I = I' \cup I''$ instead of $I = I' \times I''$. This 
definition is inadequate for permutation problems like sorting
or even planar hulls. 
}

%\subsection{Oblivious, adaptive and estimation}

%\subsection{Sorting inputs with measurable orderliness}

%It is easy to verify that the given input is sorted using $n-1$ comparisons, so we
%can clearly do better. 

\subsection{Further work}

Our algorithm is a simple variation of the standard partition sort and
for implementation, it may be better to use an appropriate modification of the 
randomized Quicksort by incorporating stability and early termination. 
The analysis of the expected running time of
$O( n \cdot H ( \frac{n_1}{n} , 
\frac{n_2}{n} \dots \frac{n_k}{n}) + n)$ bound needs to be strengthened
with inverse polynomial deviation bounds. 
Unlike many of the previous algorithms that exploit pre-sortedness,
our algorithm doesn't require any sophisticated data structures and its 
effectiveness is derived from a natural parametrization.

One interesting direction is to generalize our approach by considering 
sorted subsequences like SUS/SMS whose elements are not contiguous in the final
sorted sequence. The results for the {\it slabsort} algorithms 
in \cite{LP:94} are dependent on $k$ but converting
this to a meaningful entropy dependent bound remains open.

%The input in row 3 of Figure \ref{unsorted} is more amenable 
%to {\it merging} rather than partition sort. 
%However, proving the optimality of algorithms based on these measures of 
%pre-sortedness is unclear.
%Recently, the authors in \cite{GJKY:25} 
%showed that exponential merging 
%achieves an optimal number of comparisons in adaptive mergesort. 
%However, mergesort requires a lot of data
%movements that are not known to be optimal.
%This appears to be difficult for 
%merging based sorting algorithms, where, the running time depends on the 
%explicit decomposition of the input into SUS/SMS. although the number of
%comparisons can be bounded by this entropic measure \cite{LP:94,GJKY:25}.
The computational hardness of decomposing a 
sequence into minimal SMS or minimal entropy SUS and SMS has been noted 
\cite{BN13} which makes them impractical.
Compared to this, our modified Quicksort algorithm is particularly 
attractive since 
it does not explicitly find the $\alpha$-sorted subsequences. Closing the
gap between the upper and lower bounds for the case of $k = n - o(n)$ is
an interesting open problem that is related to some known fundamental problems.  
Another interesting open problem related to our work is to extend this to
an in-place version of our algorithm.

%The paradigm of universal optimality has an inherent dependence on the parametrization
%used by the algorithm, therefore, there may not be one perfect algorithm. For example,
%partition sort and mergesort would be effective for different families of input
%permutations. We have seen that other sorting strategies like insertion sort and heap
%sort are effective for inversion based measures. Our contribution has shown
%that the standard partition sort can exploit presortedness in a very 
%natural way. 

%\\ [0.1in]
{\bf Acknowledgement} The author is thankful to an anonymous reviewer for some
insightful technical comments on an earlier version of the draft.
The author leveraged ChatGPT Plus for improving the 
readability of some proofs and illustration of some examples. 
\bibliography{references}
\appendix
\section{Appendix}
\ignore{
We can extend this to the general case as follows.
\begin{lemma}
Let $n_1+\cdots+n_k=n$, where $n_i\ge1$, and let
$\mathcal A$ denote the permutations admitting the prescribed
$\alpha$-sorted decomposition $(n_1,\ldots,n_k)$. Let
$\mathcal B\subseteq\mathcal A$ denote those permutations in which
all prescribed subsequences are maximal. Then
\[
|\mathcal B|
\ge
\frac{1}{4^{k-1}}
\frac{n!}{n_1!\cdots n_k!}.
\]
\end{lemma}

\begin{proof}
When all $n_i\ge2$, the mergeable prescribed boundaries can be
repaired by disjoint swaps of consecutive-rank elements, and the
set of repaired boundaries gives a signature from at most
$2^{k-1}$ possibilities.

For general $n_i\ge1$, first group consecutive prescribed blocks
into packets so that every packet has total length at least two.
The grouping is performed canonically from left to right by pairing
singleton blocks with adjacent blocks. The required transformation
is carried out in two stages: first repair mergeable boundaries
between packets, and then repair prescribed boundaries internal to
the packets.

The transformation is described by a two-level signature
\[
\Sigma=(\Sigma_{\rm out},\Sigma_{\rm in}),
\]
where each component is a subset of the $k-1$ prescribed
boundaries. Hence the number of possible signatures is at most
\[
2^{k-1}\cdot2^{k-1}=4^{k-1}.
\]

Fix a maximal permutation and a two-level signature. Since the
packet construction and the order of the two stages are canonical,
and every swap is an involution, they determine at most one
candidate preimage. Thus every vertex on the maximal side of the
bipartite graph has degree at most $4^{k-1}$.

Every vertex of $\mathcal A$ is mapped to some vertex of
$\mathcal B$, and therefore
\[
|\mathcal A|\le4^{k-1}|\mathcal B|.
\]
Since
\[
|\mathcal A|=\frac{n!}{n_1!\cdots n_k!},
\]
the claimed bound follows.
\end{proof}
}%ignore
\begin{example}
Consider the prescribed ordered type
\[
(2,2,2,2,2)
\]
on \(n=10\) elements.  The corresponding consecutive rank blocks are
\[
R_1=\{1,2\},\quad
R_2=\{3,4\},\quad
R_3=\{5,6\},\quad
R_4=\{7,8\},\quad
R_5=\{9,10\}.
\]

It is convenient to work with the inverse permutation
\[
q=\pi^{-1}.
\]
The prescribed blocks form $\alpha$-sorted subsequences precisely when
the entries of \(q\) are increasing within each block.  For these
subsequences to be maximal, there must additionally be a descent at
each prescribed boundary.

Consider the blockwise increasing permutation
\[
q=
(1,2)\mid(3,4)\mid(7,8)\mid(5,6)\mid(9,10).
\]
Equivalently,
\[
q=(1,2,3,4,7,8,5,6,9,10).
\]
Each prescribed block is internally increasing.  However, the boundary
comparisons are
\[
2<3,\qquad
4<7,\qquad
8>5,\qquad
6<9.
\]
Hence the first, second, and fourth boundaries are mergeable.  The
actual maximal increasing runs are
\[
(1,2,3,4,7,8)\mid(5,6,9,10),
\]
and therefore this permutation does not have maximal run lengths
\((2,2,2,2,2)\).

The prescribed boundaries lie between the consecutive ranks
\[
2\mid3,\qquad
4\mid5,\qquad
6\mid7,\qquad
8\mid9.
\]
For every mergeable boundary, swap the two corresponding entries of
\(q\).  Thus, we swap the entries in positions \(2\) and \(3\), the
entries in positions \(4\) and \(5\), and the entries in positions
\(8\) and \(9\).  The boundary between the third and fourth blocks is
already non-mergeable and is left unchanged.

The resulting permutation is
\[
q'
=
(1,3,2,7,4,8,5,9,6,10),
\]
which, according to the prescribed blocks, can be written as
\[
q'
=
(1,3)\mid(2,7)\mid(4,8)\mid(5,9)\mid(6,10).
\]

Every prescribed block remains internally increasing:
\[
1<3,\qquad
2<7,\qquad
4<8,\qquad
5<9,\qquad
6<10.
\]
Moreover, every prescribed boundary is now a descent:
\[
3>2,\qquad
7>4,\qquad
8>5,\qquad
9>6.
\]
Consequently, the maximal increasing runs of \(q'\) have lengths
exactly
\[
(2,2,2,2,2).
\]

Since the maximal increasing runs of \(q'=\pi^{-1}\) correspond to the
maximal $\alpha$-sorted subsequences of \(\pi\), the corresponding input
permutation is
\[
\pi=(q')^{-1}
=
(1,3,2,5,7,9,4,6,8,10).
\]
Thus \(\pi\) has maximal $\alpha$-sorted subsequences of lengths
\[
(2,2,2,2,2).
\]

The signature of the transformation is the set of repaired boundaries
\[
\Sigma=\{1,2,4\}.
\]
There are four prescribed boundaries, and hence at most
\[
2^4=16
\]
possible signatures.  Given \(q'\) and \(\Sigma\), the original
permutation \(q\) is recovered by applying the same swaps, since each
swap is an involution.

Finally, the number of permutations that are increasing within the
five prescribed blocks is
\[
\frac{10!}{(2!)^5}=113400.
\]
Therefore, the signature argument implies that the number of maximal
realizations of this ordered type is at least
\[
\left\lceil
\frac{113400}{16}
\right\rceil
=
7088.
\]
\end{example}
%\begin{thebibliography}{99}  
%\end{thebibliography}
\end{document}